\documentclass[conference]{IEEEtran}

\usepackage{listings}
\usepackage{multirow}
\usepackage{graphicx}
\usepackage{xcolor}

\newcommand{\comment}[1]{}

\begin{document}

\lstset{language=C,basicstyle=\small}
\lstset{numbers=left, numberstyle=\tiny, stepnumber=1, numbersep=5pt}
\lstset{tabsize=2}
\lstset{firstnumber=1}
\lstset{frame=single}
\lstset{
  language={C},
  morekeywords={assert,uchar}
}
	
	\title{Fault Localization in Multi-Threaded C Programs using Bounded Model Checking (extended version)}
	
	\author{\IEEEauthorblockN{Erickson H. da S. Alves, Lucas C. Cordeiro, Eddie B. de Lima Filho}
		\IEEEauthorblockA{Federal University of Amazonas, Brazil\\
			E-mails: erickson.alves@indt.org.br, lucascordeiro@ufam.edu.br, eddie@ctpim.org.br}}
	
	\maketitle
	
	\begin{abstract}
		Software debugging is a very time-consuming process, which is even worse for multi-threaded programs, due to the non-deterministic behavior of thread-scheduling algorithms. However, the debugging time may be greatly reduced, if automatic methods are used for localizing faults. 
		In this study, a new method for fault localization, in multi-threaded C programs, is proposed. It transforms a multi-threaded program into a corresponding sequential one and then uses a fault-diagnosis method suitable for this type of program, in order to localize faults. The code transformation is implemented with rules and context switch information from counterexamples, which are typically generated by bounded model checkers. Experimental results show that the proposed method is effective, in such a way that sequential fault-localization methods can be extended to multi-threaded programs.
	\end{abstract}
	
	
	\IEEEpeerreviewmaketitle
	
	\section{Introduction}
	
	Recently, it has become more and more common for technology to handle various tasks in everyday life, each one with an associated complexity. Ensuring that systems work properly imply in cost reduction and even, in some areas, that lives are safe. This is what makes program debugging so worthy of attention in computer-based systems. Program debugging is a very important but time-consuming task, in software development processes, which can be divided into three steps: fault detection, fault localization, and fault correction. However, the associated debugging time can be largely reduced, if automatic methods are used for localizing faults, especially in multi-threaded programs, which are widely used in embedded system products.
	
	A number of different approaches have been introduced, in order to provide automated methods for localizing faults in applications, based on the generation of a program model that is extracted from its source code~\cite{mayer}. Those include slicing~\cite{tip}, mutation testing~\cite{offutt}, trace-based analysis ~\cite{he}, delta-debugging ~\cite{cleve}, model-based debugging~\cite{friedrich}, and model checking ~\cite{chaki}. In this paper, a fault localization method is proposed, which relies entirely on model-checking techniques. In particular, Bounded Model Checking (BMC) based on Satisfiability Modulo Theories (SMT) is used to automatically refute a safety property and consequently produce a counterexample, if the (multi-threaded) program does not satisfy a given specification. It is worth noticing that the generated counterexamples may be regarded as error traces, which contain useful information about faults, so that one can localize and correct them. 
	
	The C Bounded Model Checker (CBMC)~\cite{clarke3} and also the Efficient SMT-Based Context-Bounded Model Checker (ESBMC)~\cite{cordeiro} are both well known BMC tools, which are suitable for verifying multi-threaded C programs. Since the majority of the initiatives, in multi-threaded software verification, have focused on Java, as shown by Park, Harrold, and Vuduc~\cite{park}, this study suggests an automated fault localization method for multi-threaded C programs, using SMT-based BMC techniques. In particular, in this present work, ESBMC~\cite{cordeiro} is adopted, since it is one of the most efficient verifiers, as reported by Beyer~\cite{beyer}, and it also supports both single- and multi-threaded programs, using different SMT solvers to check for the generated verification conditions (VCs)~\cite{cordeiro3}. 
	
	Two main goals are achieved here: the evaluation of the method proposed by Griesmeyer, Staber, and Bloem~\cite{griesmeyer} and the improvement/extension of this same method, in order to support multi-threaded applications. The basic concept of extending the mentioned work~\cite{griesmeyer} consists in transforming a multi-threaded program into a corresponding sequential one, by carrying out evaluation and transformation steps, and then using that work for localizing faults. As a consequence, the found violations, in the sequential code, can show the location of the initial faults, in the original multi-threaded program.
	
	This paper is organized as follows. Section \ref{sec:relatedwork} describes the related work. Fault localization based on model checking is introduced in section \ref{sec:background}. The method proposed by Griesmeyer, Staber, and Bloem \cite{griesmeyer} is also tackled, in detail, in section \ref{sec:background}. Section \ref{sec:methodmultithreaded} is the main section of the present paper and demonstrates the proposed method for fault localization, in multi-threaded programs. Section \ref{sec:experiment} provides experimental results, analysis, and discussion. Finally, section \ref{sec:conclusion} summarizes the main results and highlights future work.
	
	\section{Related work}
	\label{sec:relatedwork}
	
	Cleve {\it et al.}~\cite{cleve} show how {\it cause transitions}, which are moments where a variable replaces another as failure cause, can locate defects in programs. Such an achievement is possible due to a comparison regarding program states of failing and passing runs; however, given that such state differences can occur all over the program run, the focus is {\it in space}, with a subset of variables that is relevant to the failure occurrence, and also {\it in time}, where {\it cause transitions} occur.
	
	Griesmeyer {\it et al.}~\cite{griesmeyer} proposed a method for localizing faults, in ANSI-C programs, by instrumenting the original code and running that new version on a model checker. Such an approach is very helpful, given that model checkers are able to identify the exact fault line; however, this work only presented a method for sequential programs~\cite{griesmeyer}.
	
	Birch {\it et al.}~\cite{birch} describe a method for fast model-based fault localization, which, given a test suite, automatically identifies a small subset of program locations, where faults exist, by using symbolic execution methods. In summary, the mentioned algorithm tries to find counterexamples that are capable of localizing faults, based on failing test cases from a test suite. The key factor to its speed is that if an execution takes longer than expected, it is pushed into a queue, for later handling, and then another execution is chosen to be run.
	
	Jones {\it et al.}~\cite{jones} showed how test information visualization can assist in fault localization. By coloring program statements that participate in the outcome of a program execution, with a test suite, it is possible to assist users to inspect code, evaluate statements involved in failures, and identify possible faults.
	
	Jose {\it et al.}~\cite{jose} reported a method to localize faults in programs, using a reduction to the Maximal Satisfiability Problem, which informs the maximum number of clauses, of a Boolean formula, that can simultaneously be satisfied by an assignment. The potential error is given by finding the maximal set of clauses that can be satisfied, in a formula generated by combining a failing program execution and a Boolean trace formula, and outputting the complement set.
	
	Tomasco {\it et al.}~\cite{tomasco} presented an approach for symbolically verifying multi-threaded programs, with shared memory and dynamic thread creation, by using a technique called Memory Unwinding (MU), which is the process of writing operations into shared memory. A code-to-code transformation from multi-threaded to sequential programs was used, following MU rules, and then checked by a sequential verification tool.
	
  The closest related work is that of Park {\it et al.}~\cite{park}, which describe a dynamic fault localization method to localize the root causes of concurrency bugs in Java programs, based on dynamic pattern detection and statistical fault localization. To the best of our knowledge, this present paper marks the first application of a fault-localization method, based on BMC techniques, to a broader range of multi-threaded C programs.
		
	\comment{This work presents a new method to localize faults in multi-threaded programs by instrumenting (similar to what Griesmeyer {\it et al.}~\cite{griesmeyer} suggested) a sequential version of the original code. However, to obtain this sequential version of the multi-threaded program, some rules and grammar are needed in order to maintain the original execution.}
	
	\section{Background}
	\label{sec:background}
	
	\subsection{Bounded Model Checking to Multi-threaded Software}
	\label{sec:bmc}
	
	The basic idea of BMC applied to multi-threaded software is to check for the negation of a given property, at a given depth~\cite{cordeiro11}. Given a reachability tree $\Upsilon = \{\nu_1,\ldots, \nu_N\}$, which represents the program unfolding for a context bound $C$ and a bound $k$, and a property $\phi$, BMC derives a VC $\psi^{\pi}_{k}$ for a given interleaving (or computation path) $\pi = \{\nu_1,\ldots, \nu_k\}$, such that $\psi^{\pi}_{k}$ is satisfiable if and only if $\phi$ has a counterexample of depth \textit{k}, which is exhibited by $\pi$.\ The VC $\psi^{\pi}_{k}$ is a quantifier-free formula in a decidable subset of first-order logic, which is checked for satisfiability by an SMT solver~\cite{Z08}.\ The model checking problem associated with SMT-based BMC, of a given $\pi$, is formulated by constructing the logical formula~\cite{cordeiro3}
	\begin{equation} \label{bounded-model-checking}
		\psi^{\pi}_{k} =
		\overbrace{I(s_{0}) \wedge R(s_{0},s_{1})\wedge\ldots\wedge
			R(s_{k-1},s_{k})
		}^{constraints}
		\wedge \overbrace{\neg \phi_{k}}^{property}.
	\end{equation}
	Here, $\phi_{k}$ represents a safety property $\phi$, in step
	$k$, $I$ is the set of initial states, and $R(s_{i},s_{i+1})$ is the
	transition relation at time steps $i$ and $i+1$, as described by the states in $\pi$ nodes.\ 
	In order to check if (\ref{bounded-model-checking}) is satisfiable or
	unsatisfiable, the SMT solver constrains some symbols, by a given
	background theory ({\it e.g.}, arithmetic restricts the
	interpretation of symbols such as $+$, $\leq$, $0$, and
	$1$)~\cite{Z08}.\ If (\ref{bounded-model-checking}) is
	satisfiable, then $\phi$ is violated and the SMT solver provides a satisfying assignment, 
	from which one can extract values of program variables to construct a counterexample, {\it i.e.}, 
	a sequence of states $s_{0}, s_{1},\ldots, s_{k}$, with $s_{0} \in S_{0}$, 
	and transition relations $R(s_{i}, s_{i+1})$, for $0 \leq i < k$.\ 
	If (\ref{bounded-model-checking}) is unsatisfiable, one can conclude that no
	error state is reachable, in length $k$ along $\pi$.\ 
	
	\comment{
		\subsection{ESBMC}
		\label{sec:esbmc}
		
		ESBMC is a context-bounded model checker based on SMT, which is used for ANSI-C/C$++$ programs~\cite{cordeiro}. ESBMC verifies sequential and multi-threaded programs and also checks for properties related to arithmetic overflow, division by zero, out-of-bounds index, pointer safety, deadlock, and data race. In ESBMC, the verification process is completely automated and does not require users to annotate programs, with pre/post-conditions.
		
		Fig.~\ref{figure:esbmc-architecture} shows the ESBMC architecture. As one can see, ESBMC converts an ANSI-C/C$++$ program into a \textit{GOTO} one, which simplifies statement representations ({\it e.g.}, replacement of \textit{while} by \textit{if} and \textit{goto} statements). Then, the resulting \textit{GOTO} program is executed symbolically by the \textit{GOTO} symex, which generates a Single Static Assignment (SSA) form that is later converted into a first-order logic formula; the latter is finally checked by an SMT solver. If a property violation is found, then a counterexample is provided by ESBMC, which assigns values to program variables, in order to reproduce the associated error.
		
		\begin{figure}[ht]
			\centering
			\includegraphics[scale=0.2]{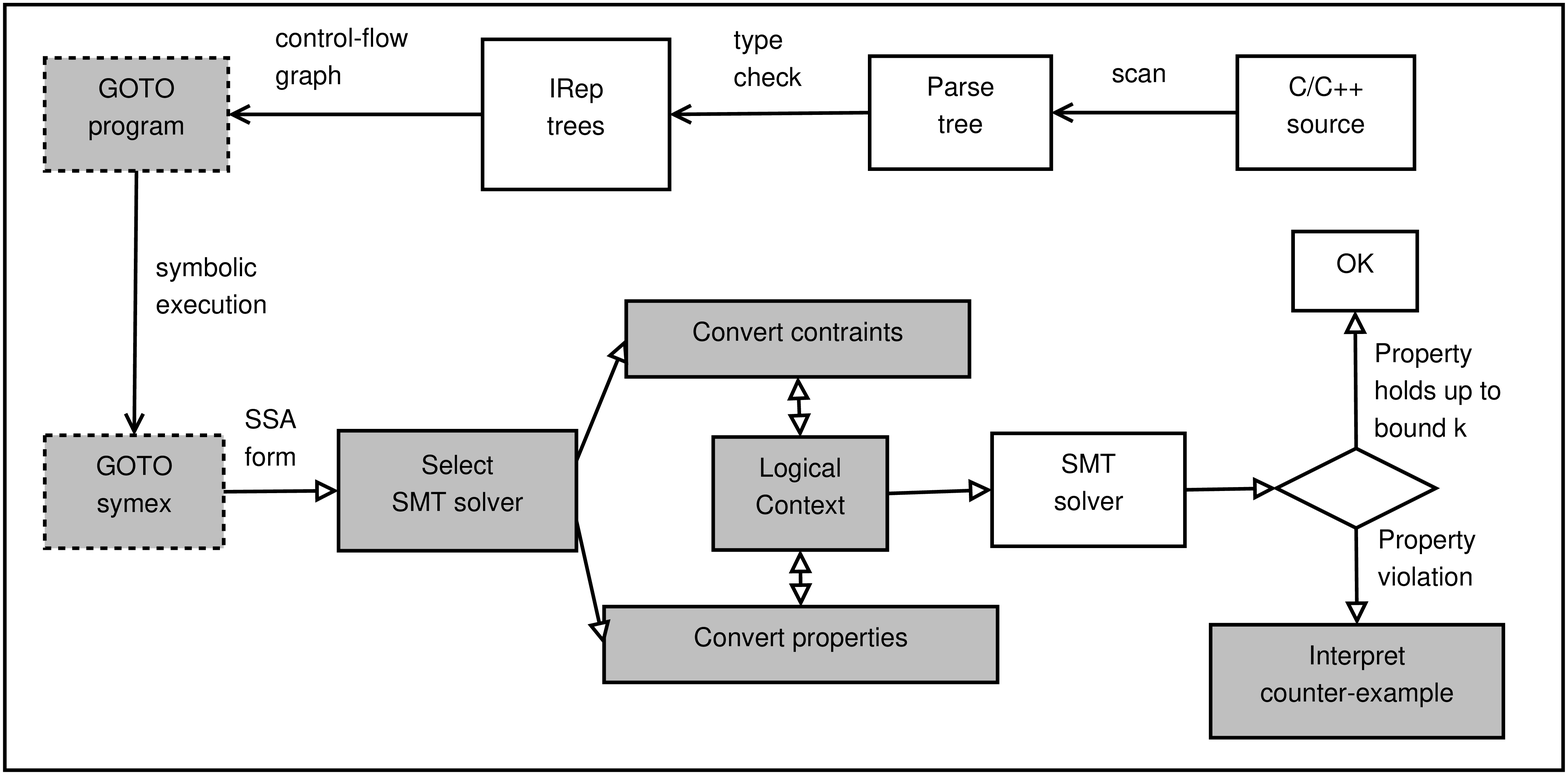}
			\caption{Overview of the ESBMC architecture.}
			\label{figure:esbmc-architecture}
		\end{figure}
	}
	
	\subsection{Fault Localization using Model Checking}
	\label{sec:faultlocalization}
	
	The most basic task regarding fault localization, in model checking, is to generate a counterexample, which is provided when a program does not satisfy a given specification. According to Clarke {\it et al.}~\cite{clarke,clarke2}, a counterexample does not solely provide information about the cause-effect relation of a given violation, but also about fault localization. However, since an enormous amount of information is presented in a counterexample, actual fault lines may not be easily identified.
	
	Several methods have been proposed, in order to localize possible fault causes, through counterexamples. An approach proposed by Ball, Naik, and Rajamani ~\cite{tball} tries isolating possible causes of counterexamples, which are generated by the SLAM model checker~\cite{tball2}. In summary, potential fault lines can be isolated by comparing transitions among obtained counterexamples and successful traces, since transitions not included in correct traces are possible causes of errors. Groce and Visser~\cite{groce} state that if a counterexample exists, a similar but successful trace can also be found, using BMC techniques. Program elements related to a given violation are implicated by the minimal differences between that counterexample and a successful trace, which is known as the Java Pathfinder approach~\cite{jpf} and can also provide execution paths that lead to error states, regarding for multi-threaded programs ({\it e.g.}, data race). The essence of the approach described by Groce {\it et al.}~\cite{groce2} is similar to the latter and uses alignment constraints to associate states, in a counterexample, with corresponding states in a successful trace, which was generated by a constraint solver. The mentioned states are abstract states over predicates, which represent concrete states in a trace. By using distance metric properties, constraints can be employed for representing program executions, and non-matching constraints that represent concrete states might lead to faults. Additionally, if a distance metric property is not satisfied, a counterexample is generated by the BMC tool~\cite{groce2}.
	
	In contrast to the transition-based and difference-based methods mentioned above, a method can directly identify possible faults by combining instrumented programs and BMC, as shown by Griesmeyer {\it et al.}~\cite{griesmeyer,griesmeyer2,griesmeyer3}, which will be further demonstrated. The approach proposed in the present paper is based on that method and consists in an extension to multi-threaded programs, that is, it tries to identify fault lines in multi-threaded programs, using BMC techniques.

	\subsection{Method demonstration}
	\label{sec:methoddemo}
	
	The method proposed by Griesmeyer {\it et al.}~\cite{griesmeyer} is based on the BMC technique, which can directly identify possible faults in programs. In particular, this method adds additional numerical variables ({\it e.g.}, \texttt{diag1, ...,diagn})) to identify a fault in a given program. Each line of a program, representing a statement \texttt{S}, is changed to a logic version of that statement. As a consequence, the value held by \texttt{S} will be either non-deterministically chosen by the BMC tool, if the value of \texttt{diag} is the same as the one representing the line related to statement \texttt{S}, or the one originally specified. 
	
	If the BMC tool identifies a \texttt{diag} value, by correcting this line in the original program, the fault can be avoided. In the case of multiple \texttt{diag} values, correcting those lines lead to a successful code execution. In order to find the full set of lines that cause a faulty behavior in a program, a new line\footnote{\texttt{assume(diag != a)} (\texttt{a} is the line number obtained in the last run)} can be added to its source code, which is then rerun in the BMC tool. This process is repeatedly executed, until no more values of \texttt{diag} are obtained ({\it i.e.}, the run succeeds)~\cite{griesmeyer2}.
	
	As an example, a simple program slightly modified from Griesmeyer {\it et al.}~\cite{griesmeyer2}, is presented in Fig.~\ref{figure:simpleprogram}. Its modified version, using the mentioned method~\cite{griesmeyer2} and ready to be run by a BMC tool, is shown in Fig.~\ref{figure:modifiedversion}. The diagnosis informed by a BMC tool is \texttt{diag == 4} and \texttt{diag == 7}, which means that changing line $4$ (to ``\texttt{a = 6}'') or line $7$ (to \texttt{if(0)}), in the original program, can result in source code that is able to successfully execute, therefore, one can note that the example below contains a single fault.

\begin{figure}[ht]
\centering
\begin{minipage}{0.45\textwidth}
\begin{lstlisting}
void main() {
  int a, b, c, d;
	if (a) {
	  a = 5;
	  b = 2;
	  c = a + b;
	  if (a % 2 == 0) {
	  int d;
	  a = d;
	  }
	  assert (c == 8);
	}
}
\end{lstlisting}
\end{minipage}
\caption{A simple ANSI-C program with a single fault.}
\label{figure:simpleprogram}
\end{figure}
	
\begin{figure}[ht]
\centering
\begin{minipage}{0.45\textwidth}
\begin{lstlisting}
int nondet();
void main() {
  int a, b, c, d;
	int diag;
	diag = nondet();
	a = 5; b = 2; c = 7;
	if (diag == 4? nondet(): a) {
	  a = (diag == 6? nondet(): 5);
	  b = (diag == 7? nondet(): 2);
	  c = (diag == 8? nondet(): (a + b));
	  if (diag == 9? nondet(): (a%2==0)) {
	    int d;
	    a = (diag == 12? nondet(): d);
	  }
	  assume(c == 8);
	}
	assert(false);
}
\end{lstlisting}
\end{minipage}
\caption{The diagnosis model of the example shown in Fig.~\ref{figure:simpleprogram}, where \textit{nondet()} represents a non-deterministic function.}
\label{figure:modifiedversion}
\end{figure}

	\section{Fault Localization in Multi-Threaded C Programs using BMC}
	\label{sec:methodmultithreaded}
	
	The proposed method, which has the goal of localizing faults in multi-threaded C programs, is based on Griesmeyer's method~\cite{griesmeyer2} and counterexamples generated by BMC tools, such as ESBMC. Its key concept is to transform a multi-threaded program into a corresponding sequential one and then apply instrumentation for identifying faults~\cite{griesmeyer2}.
	
	\subsection{Transformations from Multi- to Single-threaded Programs}
	\label{sec:sequentialtransformation}
	
\begin{figure}[ht]
		\centering
		\includegraphics[scale=0.5]{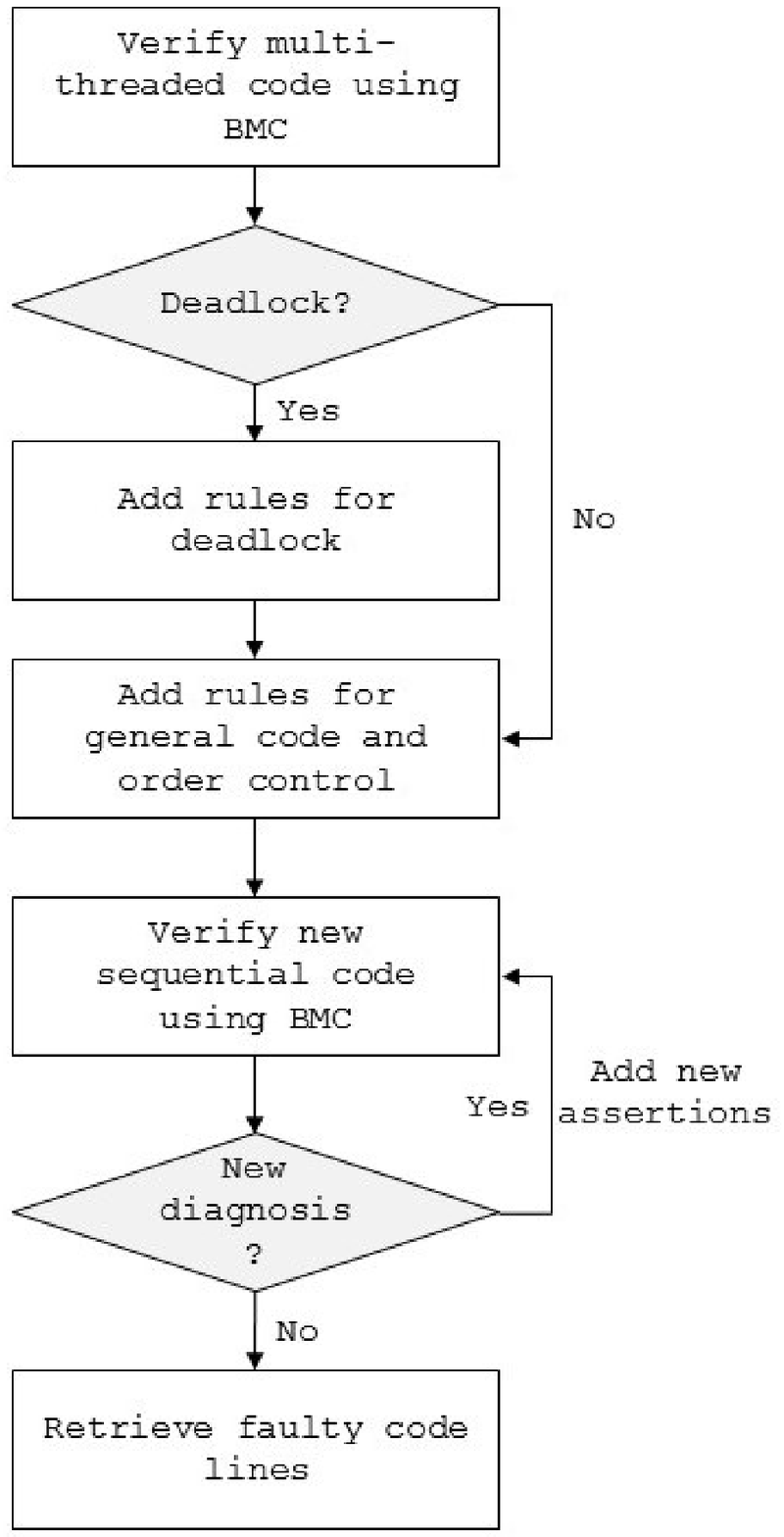}
		\caption{Proposed method for fault localization in multi-threaded software.}
		\label{figure:methodology}
	\end{figure}

	The transformation from multi-threaded programs into sequential ones can be split into four distinct steps. First, counterexamples are obtained from a BMC tool, which contain useful pieces of information related to faults. Then, the framework described below is applied, which consists in code used as fixed structure for a new sequential version, together with the use of some rules (defined later in section~\ref{sec:rules}). Following that, an original (multi-threaded) program is converted into its sequential version and, finally, order control is included into the latter, specifying the order in which threads are executed. These steps are summarized in Fig.~\ref{figure:methodology}.
	
	\comment{
		The first step consists in acquiring counterexamples of the original program using the respective BMC tool, since they provide errors (or more specifically, error pattern) and detailed threads interleavings demonstrations, including the thread in which a context switch occurs, the number of context switches, the specific code line, where a context switch occurs, and variables changes; therefore, counterexamples play an important role in the proposed method.
		
		The second step is basically to create a new file using the basic structure in the framework to organize the new sequential code so that one maintains its execution order and threads interleavings.
		
		The third step consists in using rules for code lines in the original (multi-threaded) program to be transformed to the new sequential versions, and these rules can be slightly different according to the type of the error detected, {\it i.e.}, if it is a deadlock or not.
		
		By adding order control in the forth step, statements in the new sequential program are executed in the same sequence as in the original multi-threaded program, which is guaranteed by using the detailed interleaving information from counterexamples previously obtained. In other words, the error occurred in the original program is preserved in the new sequential version.}
	
	A framework provides the same execution sequence as in the original program. It consists basically in writing each thread code inside a $case$ statement, and their execution sequence is specified in the $order$ array. Such a framework is used as the basic structure for new sequential versions of multi-threaded programs, and Fig.~\ref{figure:framework} shows how it is encoded.

\begin{figure}[ht]
\centering
\begin{minipage}{0.45\textwidth}
\begin{lstlisting}
int order[1] = {1};
int main(int argc, char *argv[]) {
	int order_index;
	for(order_index= 0; order_index < 1; 
	    order_index++) {
	  switch(order[order_index]) {
	    case 1:
    	    case 11: { ... }
    	    ...
    	    case 20: { ... }
	    break;
	    case 2:
    	    case 21: { ... }
    	    ...
    	    case 30: { ... }
	    break;
	    case 3:
	        case 31: { ... }
    	    ...
    	    case 40: { ... }
    	break;
	    ...
	    default:
	    break;
	  }
	}
	return 1;
}
\end{lstlisting}
\end{minipage}
\caption{The standard framework to localize faults in sequential code.}
\label{figure:framework}
\end{figure}

	As one can note, the mentioned framework provides new fixed positions, for each part of the original code, and Table~\ref{table:relation} shows the relation between new positions and code-fragment types, that is, it summarizes how the new sequential code is structured. In particular, global elements, global variables, header file declarations, and other types of global declarations are placed before the sequential code $main$ function. The body of its $main$ function, in the original code, is placed between the $case$ $1$ statement and its respective $break$ command, the body of the first thread is placed between $case$ $2$ and its respective $break$ command, and so on. This process is repeated until there are no more threads to be inserted into the sequential code version. Additionally, arguments passed to the original program $main$ function are all passed to the sequential version $main$ one. In cases where threads are partially executed, a context switch occurs, another thread is executed, or a previous thread continues to execute from where it stopped, the respective pieces of code are inserted into each $case$ inside the $N^{th}$ $case$ (the $N^{th}$ case represents the $N^{th}$ thread), in such a way that the execution order remains the same.
	
	\begin{table}[ht]
		\renewcommand{\arraystretch}{1.3}
		\caption{Relation between positions and codes}
		\label{table:relation}
		\centering
		\begin{tabular}{c|c}
			\hline \bfseries Code Fragment Type & \bfseries Position in the\\
			\bfseries in the Original Code & \bfseries New Sequential Code\\
			\hline global elements & before line $1$\\
			\hline main function body & between ``case $1$'' and ``break''\\
			\hline thread body $n$ & between ``case $n + 1$'' and ``break''\\
			\hline
		\end{tabular}
	\end{table}
	
	In order to maintain the same execution order found in the original program, switch order control is required. A fixed context switch order, from a counterexample of a multi-threaded program, can be copied to a new sequential one by controlling ``$case$'' and conditional statements\footnotemark, in the framework $switch$ statement. In general, adding context switch order control to the new sequential program can be divided into two steps. In order to show a simple situation for illustrating that, it is assumed that there are less than $10$ context switches in each thread ($\forall N_{ti}, N_{ti} < 10$), a counterexample, given by a BMC tool, has $N$ context switches, and from those $N$ context switches, $N_{t0}$ occur in the main function, $N_{t1}$ occur in thread $1$, $N_{t2}$ in thread $2$, and so on $(N_{t0} + ... + N_{tn} = N)$.
	
	The first step is to get information from counterexamples generated by a BMC tool, {\it i.e.}, the total number of context switches in the original program and in each thread, the order of all context switches in the entire program and also in a single thread, and the corresponding position where a context switch occurred. With such data, it is possible to add conditional statements\footnotemark[\value{footnote}] for maintaining the same execution order of the original program, so that when a line is executed, the sequential code executes the next $case$ statement, which represents the next thread in the original code.
	
	\footnotetext{$if (order[order\_index]) == X) \: \: break;$, where $X$ represents the number of the context switch}
	
	One can note that if there are iteration statements in the original multi-threaded program, for every iteration statement, a global variable named ``$loopcounter$'' is added. Besides, a statement to increment the value of “$loopcounter$” is also added to the end of each loop body. This newly added global variable is used as a condition to directly control the validity of $break$ statements, so that when a context switch occurs, inside a loop, then the value held by $loopcounter$ must also be used in the respective $break$ statement, in order to maintain the original program execution sequence.
	
	The second step consists in modifying values related to the $order$ array, in such a way that the execution order in kept, in a new sequential program. By changing lines $1$ and $4$, in Fig.~\ref{figure:framework}, according to the specific number of occurred context switches and their execution order, it is possible to guarantee the original execution order, since a $switch$ statement (line $6$) selects which piece of code (representing threads from the original program) is executed, based on $order[order\_index]$. For instance, if the execution order of the original code is thread $0$, thread $2$, and thread $1$ (note that this information was previously extracted from the counterexample), the $order$ array will hold $11$, $31$, and $21$, meaning that the first $case$ will be executed, then the third and, finally, the second one.
	
	\subsection{Code Transformation}
	\label{sec:codetransformation}
	
	\subsubsection{Grammar}
	\label{sec:grammar}
	
	Transformation rules, regarding code fragments, are needed, when code fragments are added to corresponding positions in the mentioned framework. Given that the most common faults, in multi-threaded programs, are related to data races and deadlocks~\cite{muhlenfeld}, a simple grammar, for code fragments, can be defined, w.r.t. these two faults types.
	
	Threads in C are typically implemented through the POSIX Pthreads~\cite{butenhof} standard, which defines an Application Programming Interface (API) for creating and handling threads. POSIX threads are available in a library, called $pthread$, which is used in UNIX operating systems. Therefore, two groups are created: one regarding {\it pthread} non-related code fragments, which is group $non-pthread$, and another for {\it pthread} related code fragments, called group $pthread$.
	
	\subsubsection{Rules}
	\label{sec:rules}
	
	The rules used to transform code fragments, in the original multi-threaded program, are shown in Table \ref{table:rules}. Note that such transformations rely on counterexamples generated by a BMC tool. Additionally, different threads are simulated by different \texttt{case} statements, since the $main$ function is in the first $case$ statement, the first executed thread is in the second $case$ statement, and so on, as already mentioned.
	
	\begin{table}[ht]
		\renewcommand{\arraystretch}{1.3}
		\caption{Rules to transform multi-threaded programs}
		\label{table:rules}
		\centering
		\begin{tabular}{c|c|c|c}
			\hline \bfseries Group & \bfseries Code fragment & \bfseries No deadlock & \bfseries Deadlock\\
			\hline \multirow{3}{*}{1}&Variable declaration&No changes&No changes\\ &Expression&Unwind&Unwind\\&Statement&No changes&No changes\\
			\hline \multirow{11}{*}{2}&pthread\_t&$\epsilon$&$\epsilon$\\
			&pthread\_attr\_t&$\epsilon$&$\epsilon$\\
			&pthread\_cond\_attr\_t&$\epsilon$&$\epsilon$\\
			&pthread\_create&$\epsilon$&$\epsilon$\\
			&pthread\_join&$\epsilon$&$\epsilon$\\
			&pthread\_exit&$\epsilon$&$\epsilon$\\
			&pthread\_mutex\_t&$\epsilon$&Integer variable is declared\\
			&pthread\_mutex\_lock&$\epsilon$&1 is assigned to variable\\
			&pthread\_mutex\_unlock&$\epsilon$&0 is assigned to variable\\
			&pthread\_cond\_t&$\epsilon$&Integer variable is declared\\
			&pthread\_cond\_init&$\epsilon$&0 is assigned to variable\\
			&pthread\_cond\_wait&$\epsilon$&1 is assigned to variable\\
			&pthread\_cond\_signal&$\epsilon$&0 is assigned to variable\\
			
			\hline
		\end{tabular}
	\end{table}
	
	In Table~\ref{table:rules}, $\epsilon$ stands for the removal of the respective statement in the new sequential version. When a deadlock is returned, by the BMC tool, one needs to add an integer variable for simulating the {\it pthread\_mutex\_t} and/or {\it pthread\_cond\_t} variables. Finally, the $unwind$ process for expressions, in Table~\ref{table:rules}, consists in removing the original function and directly writing this piece of code, {\it e.g.}, if the value returned by function $f$ is assigned to variable $i$, when it is called with argument $a$ ($i=f(a)$), such call is removed and replaced by the actual calculation. This process is described in Figures~\ref{figure:originalcodefragment} and~\ref{figure:transformedcodefragment}.

\begin{figure}[ht]
\centering
\begin{minipage}{0.45\textwidth}
\begin{lstlisting}
int f(int m) {
	int b;
	b = m;
	return m;
}
...
int i;
i = f(a);
\end{lstlisting}
\end{minipage}
\caption{Original code fragment.}
\label{figure:originalcodefragment}
\end{figure}

\begin{figure}[ht]
\centering
\begin{minipage}{0.45\textwidth}
\begin{lstlisting}	
int i;
{
	int m = a;
	int b;
	b = m;
	i = m;
}
\end{lstlisting}
\end{minipage}
\caption{Transformed code fragment from Fig.~\ref{figure:originalcodefragment}.}
\label{figure:transformedcodefragment}
\end{figure}

	In addition, if an error detected by the chosen BMC tool is a deadlock, then rules in groups $non-pthread$ and $pthread$ are applied to create the sequential version of the original program; otherwise, only rules in group $non-pthread$ are used.
	
	\section{Experimental Evaluation}
	\label{sec:experiment}
	This section is split into two parts. The experimental setup is described in section~\ref{sec:setup}, while section~\ref{sec:results} presents
	results with the proposed method. In particular, the correctness and also the performance of the proposed method are verified using standard multi-threaded programs, which contain typical {\it pthread} functions ({\it e.g.}, conditional waiting, mutex, and join).
	
	\subsection{Experimental Setup}
	\label{sec:setup}
	
	In order to verify and validate the proposed method, ESBMC v$1$.$24$.$1$ with SMT solver Boolector~\cite{brummayer} was used.
	All experiments were conducted on an otherwise idle Intel Core i$7$ -- $4500$ $1$.$8$Ghz processor, with $8$ GB of RAM and running Fedora $21$ $64$-bits operating system.
	
	The benchmarks in Table \ref{table:results} are the same used when evaluating ESBMC for multi-threaded C programs~\cite{cordeiro11}. {\it $account\_bad.c$} is a program that represents basic operations in bank accounts: deposit, withdraw, and current balance, with a mutex to control them. {\it $arithmetic\_prog\_bad.c$} is a basic producer and consumer program, using mutex and conditional variables for synchronizing operations. {\it $carter\_bad.c$} is a program extracted from a database application, which uses mutex to synchronize threads. {\it $circular\_buffer\_bad.c$} simulates a buffer, using shared variables to synchronize receive and send operations. {\it $lazy01\_bad.c$} uses a mutex to control summation operations over a shared variable and then checks its value. {\it $queue\_bad.c$} is a program simulating a data-queue structure. {\it $sync01\_bad.c$} and {\it $sync02\_bad.c$} are producer and consumer programs: the former never consumes data and the latter initializes a shared variable with some (arbitrary) data. {\it $token\_ring\_bad.c$} propagates values through shared variables and checks whether they are equivalent, through different threads. {\it $twostage\_bad.c$} simulates a great number of threads running simultaneously and, finally, {\it $wronglock\_bad.c$} simulates a large number of producer threads and the propagation of their respective values, to other threads.
	
	The experimental evaluation procedure can be split into three different steps. First, it is necessary to identify which group (see section \ref{sec:codetransformation}) a given benchmark belongs to and, in order to have this information, one needs to execute a specific command line\footnotemark, in ESBMC. If the result given by ESBMC is \textit{verification failed}, then the benchmark belongs to group $pthread$; otherwise, it belongs to group $non-pthread$. In the second step, it is necessary to add context-switch numbers through the method presented in~\ref{sec:sequentialtransformation}, which is achieved by removing the \emph{\tt --deadlock-check} option in the issued command line\footnotemark[\value{footnote}]. In the third step, the original program is transformed into a sequential one, with the information obtained from steps $1$ and $2$, by applying the rules in section~\ref{sec:rules} and the method proposed by Griesmeyer {\it et al.}~\cite{griesmeyer}.
	
	Finally, the sequential version of the program can be verified in ESBMC, using a command line\footnotemark[\value{footnote}] without the \emph{\tt --deadlock-check} option, changing the specified file, and applying the same strategy demonstrated in section~\ref{sec:methoddemo}.
	
	\footnotetext{\texttt{ esbmc} {\tt --no-bounds-check} {\tt --no-pointer-check} {\tt --no-div-by-zero-check} {\tt --no-slice} {\tt --deadlock-check} {\tt --boolector} {\tt\textless file\textgreater}}
	
	\subsection{Experimental Results}
	\label{sec:results}
	
	Table \ref{table:results} summarizes the experimental results. \textbf{F} describes the name of the benchmark, \textbf{D} identifies whether a deadlock occurred (if its value is \texttt{1}), \textbf{FE} is the amount of errors found during the fault localization process, that is, the number of different \texttt{diag} values retrieved by ESBMC, \textbf{AE} is the number of actual errors, \textbf{R} stands for the actual result (\texttt{1} if the information retrieved by ESBMC is helpful), and, Finally, \textbf{VT} is the time that ESBMC took to verify the benchmark. The question mark is used to identify tests from which no information was retrieved, due to system limitations.
	
	\begin{table}[ht]
		\renewcommand{\arraystretch}{1.3}
		\caption{Experiment results}
		\label{table:results}
		\centering
		\begin{tabular}{c|c|c|c|c}
			\hline \bfseries \bfseries F & \bfseries D & \bfseries FE/AE & \bfseries VT & \bfseries R\\
			
			\hline {\it account\_bad.c} & $0$ & $3/3$ & $0$.$102$ & $1$\\
			
			\hline {\it arithmetic\_prog\_bad.c} & $1$ & $2/2$ & $0$.$130$ & $1$\\
			
			\hline {\it carter\_bad.c} & $?$ & $?$ & $\infty$ & $?$\\
			
			\hline {\it circular\_buffer\_bad.c} & $0$ & $7/7$ & $0$.$227$ & $1$\\
			
			\hline {\it lazy01\_bad.c} & $1$ & $4/4$ & $0$.$125$ & $1$\\
			
			\hline {\it queue\_bad.c} & $0$ & $4/4$ & $0$.$934$ & $1$\\
			
			\hline {\it sync01\_bad.c} & $1$ & $1/0$ & $0$.$451$ & $0$\\
			
			\hline {\it sync02\_bad.c} & $1$ & $2/2$ & $0$.$116$ & $1$\\
			
			\hline {\it token\_ring\_bad.c} & $0$ & $1/0$ & $0$.$101$ & $0$\\
			
			\hline {\it twostage\_bad.c} & $?$ & $?$ & $\infty$ & $?$\\
			
			\hline {\it wronglock\_bad.c} & $?$ & $?$ & $\infty$ & $?$\\
			
			\hline
		\end{tabular}
	\end{table}
	
	The verification of {\it account\_bad.c} presented $3$ different \texttt{diag} values, which are in different parts of the code; however, they ultimately identified the actual fault in the original code, which was a bad assertion.
	
	The $7$ diagnosed values regarding {\it circular\_buffer\_bad.c} led to a bad assertion in the program, which is related to a loop. This way, the \texttt{diag} values indicate this loop.
	
	When checking {\it arithmetic\_prog\_bad.c}, the proposed methodology informed $2$ different \texttt{diag} values, which address a loop in thread $2$ of this program, meaning that the fault is in that specific loop. 
	
	The analysis of both {\it lazy01\_bad.c} and {\it queue\_bad.c} presented $4$ errors. In the former, ESBMC indicated that the faults lie on the code part where its shared variable is used, which led to a bad assertion. In the latter, the identified faults are related to flags providing access control to a shared variable and a loop, where they are changed, that is, the problem lies again on bad handling.
	
	{\it sync02\_bad.c} presented $2$ different values, related to a consumer thread in the original program, whose lines are related to a deadlock present in this benchmark. 
	
	Although {\it sync01\_bad.c} and {\it token\_ring\_bad.c} presented no errors, both were diagnosed with one fault. Indeed, ESBMC found a \texttt{diag} with value $0$, which is particularly odd, since there is no line $0$. Besides, even after adding an assert, ESBMC still diagnoses $0$. Indeed, both have synchronization problems and the proposed method was unable to provide useful information.
	
	The proposed methodology was not able to verify benchmarks {\it carter\_bad.c}, {\it twostage\_bad.c}, and {\it wronglock\_bad.c}, since there was not enough memory while ESBMC checked for deadlocks. This probably occurred due to the great number of threads (in case of {\it twostage\_bad.c}, and {\it wronglock\_bad.c}) or due to a very large set of data variables ({\it carter01\_bad.c}).
	
	According to the results shown in Table~\ref{table:results}, one can note that the proposed methodology was able to find faults (useful information) in $6$ out of $11$ benchmarks, which amounts to $54$.$55$\%. Note that benchmarks whose verification failed and, consequently, from which no counterexample was extracted, are also included into this evaluation. The methodology itself showed to be useful in diagnosing data race violations, since most of the used benchmarks presented a fault related to that problem. However, the proposed method needs to be improved, in order to verify deadlocks in a more efficient way, and loop transformations also need a significant work, so that threads interleaving inside loops can be better represented. 
	
	Regarding benchmarks in which no useful information was obtained, that leads to the conclusion that improved grammar and rules are needed, in order to localize faults. Apart from that, the experimental results showed the feasibility of the proposed methodology for localizing violations, in multi-threaded C programs, since ESBMC is able to provide helpful diagnosis information regarding potential faults.
	
	\section{Conclusion}
	\label{sec:conclusion}
	
	A novel method for localizing faults in multi-threaded C programs, using code transformation and BMC techniques, was proposed. It is based on the approach introduced by Griesmeyer {\it et al.}~\cite{griesmeyer} and an extension specific to handle multi-threaded programs, which is useful for embedded systems. 
	
	The experimental results revealed the performance of the proposed methodology, when localizing faults in standard multi-threaded C benchmarks. In particular, it was able to identify potential faults in multi-threaded software, in $54$.$55$\% of the chosen benchmarks. Besides, this number may change to $75$\%, if only the ones able to be verified are considered, {\it i.e.}, those where counterexamples are provided by the BMC tool (see column \textbf{VT} in Table \ref{table:results}). 
	
	As future work, new rules for code transformation and also an improved grammar will be developed, in order to increase the methodology accuracy. Additionally, an {\it Eclipse} plug-in will be developed for automating the fault diagnosis process, during development.

\end{document}